\documentclass[]{rnaastex}

\usepackage{graphicx}
\usepackage[suffix=]{epstopdf}
\usepackage{natbib}
\usepackage{amsmath}
\usepackage{xspace}

\usepackage{hyperref}

\begin{document}

%% -----------------------------------------------
%% Editing

%\newcommand\independent{\protect\mathpalette{\protect\independenT}{\perp}}
%\def\independenT#1#2{\mathrel{\rlap{$#1#2$}\mkern2mu{#1#2}}}

\newcommand{\kepler}{\emph{Kepler}}

\newcommand{\python}{{\tt PYTHON}}
\newcommand{\phasma}{{\tt phasma}}
\newcommand{\forecaster}{{\tt forecaster}}
\newcommand{\emcee}{{\tt emcee}}
\newcommand{\scipy}{{\tt scipy}}
\newcommand{\astropy}{{\tt astropy}}
\newcommand{\sinc}{\mathrm{sinc}}

\title{A Periodogram of Every \kepler\ Target and a Common Artifact at ${\sim}80$\,Minutes}

%% Note that the corresponding author command and emails has to come
%% before everything else. Also place all the emails in the \email
%% command instead of using multiple \email calls.
\correspondingauthor{David Kipping}
\email{dkipping@astro.columbia.edu}

\author{David Kipping}
\affiliation{Department of Astronomy, Columbia University, 550 W 120th Street, New York NY}

%% Note that RNAAS manuscripts DO NOT have abstracts.

%% Start the main body of the article. If no sections in the 
%% research note leave the \section call blank to make the title.
\section{}

Studying photometric time series in the frequency domain can serve as a means
of detecting rotational modulations (e.g. \citealt{reinhold:2013,
neilsen:2013}), measuring asteroseismic modes (e.g. \citealt{chaplin:2014}) and
even detecting short-period transiting planets \citep{sanchis:2013}. To our
knowledge, there is no prior archive of \kepler\ power spectra and so we
present one here to aid the community in searching for such effects.

We downloaded the long-cadence (LC) DR25 \kepler\ PDC photometric time series
for every KIC appearing in the \citet{mathur:2017} catalog (196,845), of which
we were able to acquire the photometry for 196,791 from MAST. Photometric
points with an error flag other than zero were discarded, as were any data
occurring in Q0. Missing cadences were then filled using a spline interpolation
to form a continuous time series for each star. We then removed 3\,$\sigma$
outliers (where $\sigma$ is defined as 1.4286 multiplied by the median absolute
deviation of the spline fit residuals) against an 11-point moving median and
again replaced missing points using spline interpolation again.

Next, on each quarter independently, we computed a periodogram as the square
magnitude of the discrete Fourier transform of the photometric data (without
any weighting applied). To increase the sensitivity of the periodograms, at
the expense of some resolution, we used Welch's method \citep{smith:1999} where
the partitions were set to 1000 points in length overlapping by 500 points
each. We further apply a smoothing window to reduce ripple in the frequency
domain, in our case using the Nuttall Window and finally logged the resulting
powers. Each of these 2,594,616 Fourier transforms is made available at 
at Columbia's Academic Commons
\dataset[(DOI: 10.7916/D8RR3FCW)]{https://doi.org/10.7916/D8RR3FCW}.

The maximum frequency considered corresponds to a periodicity of twice the LC
cadence, or 24.47\,cycles per day. The lowest frequency was 0.049 cycles per 
day (corresponding to a periodicity of 20.4\,days), and this also defined the
frequency step-size used, giving a total of 500 points per periodogram evenly
spaced in frequency.

In order to investigate the possibility of global instrumental artifacts, we
elected to combine periodograms together. To do this, we first normalize the
spectra to a common scaling due to the very dynamic range in powers observed.
The normalization was chosen such that the median power from 1.5 to 1.0
hours, the highest frequency range, was set to unity. We then computed the
median and upper/lower 1\,$\sigma$ quantiles of the combined periodograms at
each frequency.

\begin{figure*}
\begin{center}
\includegraphics[width=16.0cm,angle=0,clip=true]{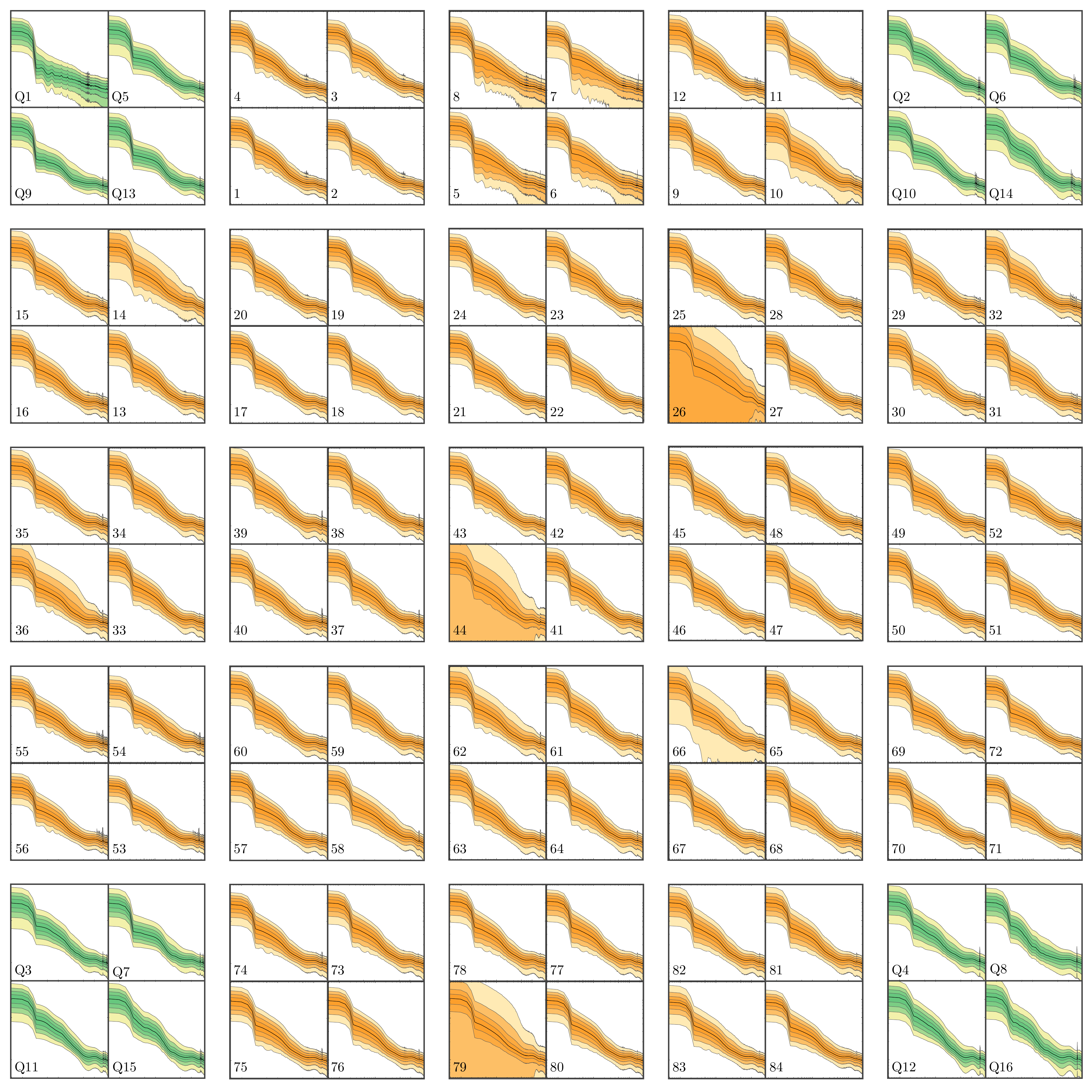}
\caption{\emph{
Black numbered panels show each of \kepler's 84 channels with a channel-averaged
periodogram (across all quarters) depicted in each box. The continuous
lines in each periodogram represent the $\pm 2$\,$\sigma$ quantiles around
the median, in 0.5\,$\sigma$ steps. Corner panels, numbered ``Q'', depict the
quarter-averaged periodograms. Full versions of each figure and the data behind each
are made available at \dataset[(DOI: 10.7916/D8RR3FCW)]{https://doi.org/10.7916/D8RR3FCW}. 
}}
\label{fig:1}
\end{center}
\end{figure*}

Channel-averaged power spectra are observed to universally peak at the lowest
frequencies (see Figure~\ref{fig:1}), as expected due to long-term trends
dominating. Moving into the higher frequencies, the power decreases for all
channels until around 3.5\,hours, after which a slight and broad excess is
observed, centered at around 2\,hours. Close to this broad and very common
bump, a small but sharp excess power is also observed for the majority of the
channels at $83.0 \pm 2.5$\,minutes (2.8 cadences), leading to a $\sim2$\%
increase in the median power. These two features may be related, as our best
estimate for the peak of of the broad bump is 136\,mins, beginning at 190\,mins
and so if symmetric in periodicity the feature would end at 82\,minutes,
precisely where the sharp feature is observed. The 82\,minute feature
corresponds to a frequency of $\sim 200$\,$\mu$Hz, which is approximately the
peak asteroseismic frequency of a $\log g \sim 3$ evolved star
\citep{brown:1991} and thus may be represent a potential source of
contamination for \kepler\ asteroseismology. We also note that quarter-averaged
periodograms tend to be self-similar every four quarters (i.e. one full rotation).

\acknowledgments

DMK is supported by the Alfred P. Sloan Foundation.

%%%%%%%%%%%%%%%%%

\end{document}